# Clustering-based accelerometer measures to model relationships between physical activity and key outcomes

Hyatt Moore IV[1], Thomas N. Robinson[2], Alexandria Jensen[1], Fatma Gunturkun[1], K. Farish Haydel[3], Kristopher I Kapphahn[1], and Manisha Desai[1]

[1]Quantitative Science Unit, Stanford University, Stanford, CA, USA
[2]Stanford Solutions Science Lab and Division of General Pediatrics, Department of Pediatrics and Stanford Prevention Research Center, Department of Medicine, Stanford University, Palo Alto, CA, USA
[3]Stanford Solutions Science Lab and Division of General Pediatrics, Department of Pediatrics, Stanford University, Palo Alto, CA, USA

## ABSTRACT

Accelerometers produce enormous amounts of data. Research that incorporates such data often involves a derived summary metric to describe physical activity. Traditional metrics have often ignored the temporal nature of the data. We build on previous work that applies unsupervised machine learning techniques to describe physical activity patterns over time. Specifically, we evaluate a summary measure of accelerometer data derived from unsupervised clustering in a regression framework through comparisons with other traditional measures: duration of time spent in different activity intensity states, Time Active Mean (TAM), Time Active Variability (TAV), Activity Intensity Mean (AIM), and Activity Intensity Variability (AIV) using data from 268 children participating in the Stanford GOALS trial. The proportion of variation explained by the new measure was comparable to that of traditional measures across regressions of three pre-specified clinical outcomes (waist circumference, fasting insulin levels, and fasting triglyceride levels). For example, cluster membership explained 25%, 11%, and 6% of the variation in waist circumference, fasting insulin levels, and fasting triglyceride levels whereas TAM explained 25%, 10%, and 6% for these same outcomes. Importantly, however, there are challenges when regressing an outcome on a variable derived from unsupervised machine learning techniques, particularly regarding replicability. This includes the processing involved in deriving the variable as well as the machine learning approach itself. While these remain open topics to resolve, our findings demonstrate the promise of a new summary measure that enables addressing questions involving a temporal component that other traditional summary metrics do not reflect.

## INTRODUCTION

### *Measures of physical activity*

Accelerometers have been increasingly used to measure physical activity objectively (Troiano et al., 2014). The richness of accelerometer data presents both challenges and opportunities. Modern accelerometers measure acceleration in three dimensions at a high frequency (e.g., 40 Hz or more) and, often, for long periods of time (e.g., 10 days). How to process and summarize high-resolution accelerometer data is an open topic. Accelerometer-based measures of physical activity may convert the accelerometer data to describe the time spent in activity states such as “percent time spent in a sedentary state.” Others summarize the data as mean ***counts*** per minute (CPM), where count – a unitless measure developed by ActiGraph (Neishabouri et al., 2022) – reflects the level of intensity of activity at a given epoch or window of time (e.g., a typical epoch may cover a 15-second interval) (Arigo et al., 2020; Banda et al., 2016; Brailey et al., 2022; Bruijns et al., 2020; Leroux et al., 2019; Romanzini et al., 2014; Rowlands et al., 2014; Santos-Lozano et al., 2012; Zhang et al., 2012). Several studies have demonstrated CPM to correlate with mechanical loading and activity (Romanzini et al., 2014; Rowlands et al., 2014; Santos-Lozano et al., 2012; Zhang et al., 2012). Any measure that summarizes activity based on the raw accelerometer data requires some processing. For example, the count is obtained by filtering and then summing the raw acceleration over consecutive, non-overlapping epochs (ActiGraph LLC, Pensacola, FL, USA). Other common approaches, termed ***cut-point*** methods, classify activity intensity using predetermined thresholds based on accelerometer signals (Romanzini et al., 2014; Bammann et al., 2021; Colley & Tremblay, 2011; Evenson et al., 2017; Fraysse et al., 2021; Kim et al., 2012; see Kim et al., 2012 for systematic review of cut-points in youth). Processing involves defining epochs, transforming raw data, and specifying cut-points to classify intensity. **Figure 1** illustrates raw tri-axial accelerations, in units of gravity, over a 24-hour period from a child participating in the Stanford GOALS trial (Robinson et al., 2013, 2021) along with its corresponding count values, CPM, and activity categories using Romanzini's cut-point approach.

**Figure 1.** Overview of accelerometer-based activity visualization of a Stanford GOALS trial participant for a 24-hour period: (1) Tri-axial acceleration signals and their vector magnitude (VM); (2) Standard deviations of acceleration signals and their mean; (3) VM of count signal at 1-second intervals; (4) VM of count signal at 1-minute intervals (counts per minute, CPM); (5) Romanzini activity categories at 1-minute intervals; (6) Mean CPM of VM count signal in 10-minute frames; (7) Romanzini activity categories at 10-minute intervals based on mean CPM; (8) Romanzini activity categories at 10-minute intervals based on the mode of 1-minute intervals.

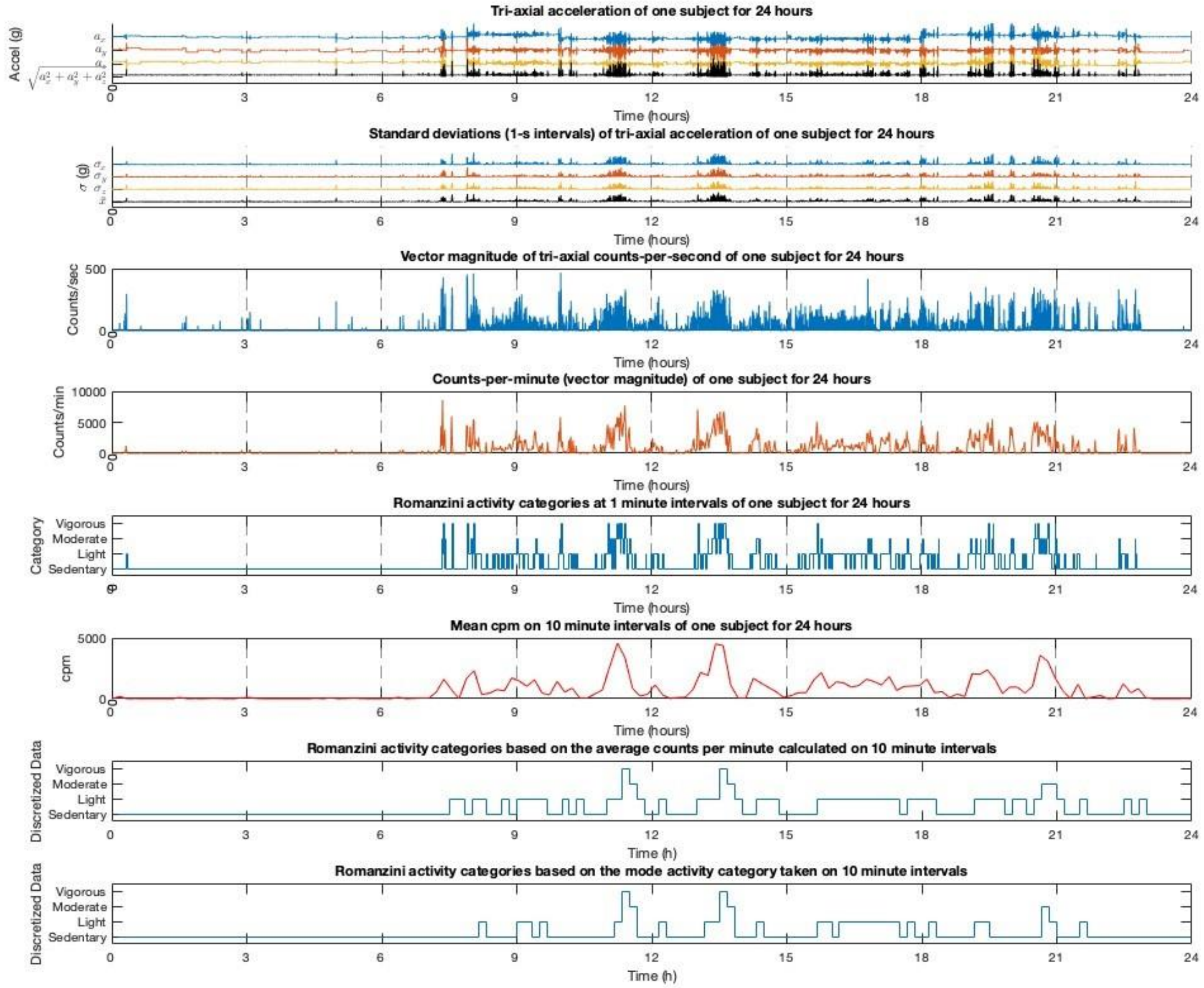

The initially proprietary nature of "count" data previously garnered some criticism by those who viewed the lack of standardization and transparency as a limitation. In their comprehensive investigation, Bai et al. addressed such limitations of many existing approaches that rely on counts (Bai et al., 2014). They proposed several new accelerometer-based summary measures in a study involving 34 older adults who wore tri-axial accelerometers for five consecutive days of normal activity and during standardized activities under laboratory observation. The daily summary measures — Time Active Mean (TAM), Time Active Variability (TAV), Activity Intensity Mean (AIM), and Activity Intensity Variability (AIV) — were developed through a transparent and comprehensive data processing pipeline. This process relies on the standard deviation of the raw signals, such as gravity, from the accelerometers. An example of this derivation can be seen in the second row of **Figure 1**. TAM represents the average duration during which activity is distinguishable from rest for each day of monitoring, while TAV calculates the variability in the duration of time when activity is distinguishable from rest for each day. AIM denotes the average amplitude of activity relative to rest, reflecting the overall intensity of the physical activities performed by an individual. AIV measures the variability in the amplitude of activity relative to rest across the days of monitoring. These four summary measures, derived from Time Active (TA) and Activity Intensity (AI) concepts, provide a more transparent and standardized approach to the analysis of raw accelerometry data but also require a cut-point to distinguish "activity" from "rest" for the sample under investigation and the accelerometer equipment used, as described in the methods section.

While these more recent methods offer greater transparency, there is still an opportunity to leverage the richness of the data to address certain types of research questions in new and innovative ways. For example, our team has been interested in questions that include the timing and patterns of physical activity that traditionally used measures may not reflect, such as whether individuals with the same amount of cumulative daily exercise who work out in the evening have more favorable cardiovascular health than those who work out in the morning. The summary measures described above do not address the pattern of physical activity signals across the day. In this work, we formally propose an alternative summary measure that leverages daily physical activity patterns preserving key temporal aspects of the accelerometer data. We evaluate the performance of the measure relative to traditional measures in a regression framework by examining the proportion of variation explained in three pre-specified clinical outcomes.

We are not the first to apply unsupervised machine learning to accelerometer data to gain insights into physical activity patterns. In a systematic review by Jones et al., 13 papers were identified that applied unsupervised machine learning to accelerometer data (Jones et al., 2021), and researchers have continued in this direction. For example, Narwin and others describe the need for measures that capture the temporal nature of intensively measured physical activity and demonstrate the diversity of physical activity patterns observed in a small data set of 42 healthy individuals (Narwin et al., 2024a). Several studies have further associated patterns – discovered by machine learning techniques – with health outcomes such as obesity, cardiovascular disease (CVD), and mental health conditions within a regression framework (Aqeel et al., 2021; Nawrin et al., 2024b; Niemelä et al., 2019; Smagula et al., 2015, 2018, 2022). These clustering-based methods provide a more granular perspective of physical activity patterns by capturing variations in the timing and intensity of activity throughout the day, rather than solely summarizing total physical activity volume. For instance, Niemelä et al. (2019) identified four temporal physical activity clusters in midlife adults and found that these patterns were significantly related to CVD risk. Similarly, Nawrin et al. (2024b) and Smagula et al. (2022) highlighted the role of physical activity timing in metabolic health and depression risk, reinforcing the idea that physical activity patterns—not just duration or intensity—may influence key health outcomes. These studies have demonstrated the promise of unsupervised machine learning approaches for gaining insight into physical activity patterns and their role in health.

There are methodological challenges, however, with leveraging a summary measure derived from unsupervised machine learning in a regression context. For example, the systematic review by Jones and authors emphasized the lack of consensus on approach and feature selection criteria making comparisons across studies difficult. The authors called for standardization of feature selection and more transparent reporting standards to improve reproducibility in the research community (Jones et al., 2021). Our study contributes to these principles by formally evaluating cluster membership as a physical activity summary measure – relative to other traditional metrics – within a regression framework when the interest is in relating physical activity patterns to a clinical outcome. We assess how one set of cluster-derived physical activity patterns compares with the existing summary measures proposed by Romanzini et al. and Bai et al., in a regression framework by focusing on the proportion of variance explained. Importantly, we additionally demonstrate the sensitivity of findings to data processing methods when deriving cluster membership when relating physical activity patterns to clinical outcome. Using data from the Stanford GOALS trial (Robinson et al., 2013, 2021), our comparative analysis highlights conditions under which pattern-based physical activity measures offer value, particularly for research explicitly addressing temporal aspects of physical activity.

## METHODS

### Stanford GOALS

Motivation for physical activity summary measures comes from the Stanford GOALS trial. Stanford GOALS was a National Institutes of Health-funded, 3-year, community-based randomized controlled trial comparing strategies for weight control among 268, 7-11-year old children with overweight or obesity from low-income, Latino (98%) families. Children had to be at or above the 85th percentile of the Center for Disease Control and Prevention's (CDC) growth charts for their age and sex to participate (Robinson et al., 2021). The baseline sample used in the current analysis consisted of 121 males and 147 females with an average age of 9.53 years and body mass index (BMI) of 25.01 (see **Table 1**).

**Table 1.** Measures from Stanford GOALS trial participants taken at baseline. Mean values shown (with standard deviation). Waist circumference, fasting insulin levels, and fasting triglycerides levels were selected for our analysis measures.

| | Male (n=121) | Female (n=147) | All (n=268) |
|---|---|---|---|
| Age (years) | 9.49 (1.50) | 9.57 (1.43) | 9.53 (1.46) |
| Height (cm) | 139.36 (9.83) | 139.08 (9.67) | 139.20 (9.73) |
| Weight (kg) | 49.64 (12.53) | 48.81 (12.99) | 49.18 (12.77) |
| Waist (cm) | 85.31 (10.79) | 85.21 (11.03) | 85.25 (10.90) |
| BMI | 25.20 (3.79) | 24.85 (4.08) | 25.01 (3.95) |
| Insulin | 13.79 (8.86) | 17.45 (12.54) | 15.80 (11.16) |
| Triglycerides | 90.00 (53.97) | 105.1 (56.46) | 98.28 (55.75) |

Participants were provided with ActiGraph (Pensacola, FL) GTX3+ ambulatory monitors[1] and instructed to wear them on their hip, continually, for a period of at least 7 days, except when bathing or swimming. The devices measured acceleration (gravity) on three perpendicular axes at a frequency of 40 Hz. Upon completion of the baseline study period, the devices were returned, and the data were exported to comma separated value (.csv) files using ActiGraph's companion software, ActiLife, which derives additional *count* measures from the raw data. These include counts for each axis of acceleration (x, y, and z) as well as their vector magnitude (VM, defined as $\sqrt{x^2 + y^2 + z^2}$).

[1] http://www.actigraph.nl/en/product/7/gt3x.html

## Physical activity data

In this study, both count and raw gravity measures from Stanford GOALS accelerometers were independently utilized and processed. The count data, obtained for the x, y, and z-axes and their VM, were exported in 1-minute, non-overlapping epochs. Likewise, the standard deviations of the raw acceleration signals were calculated using 1-sec, non-overlapping intervals and subsequently exported. This consisted of a total of 2,322 days (24 hours) from 268 boys and girls for each axis for both gravity and count based starting points. A maximum of seven days were allowed per person which brought the total to 1,850 days.

### *Data quality*

For this exercise, we only utilized completely observed data. That is, to keep our focus on the promise of a newly proposed summary metric, we excluded days where the device was not worn or functioning properly between 7:00 AM and 11:00 PM. We employed Choi's method to identify occurrences of non-wear in the count data and utilized custom code to detect periods of device malfunction in the raw gravity signal along the x, y, and z-axes[2] (Choi et al., 2011). Next, we assessed the number of days each participant could contribute, based on daily start and stop times and the presence of non-wear or malfunction indicators within that timeframe. Each day’s analysis was ultimately confined to the hours between 7:00 AM and 11:00 PM. Further details can be found in supplementary **Table S1**.

## Accelerometer Summary Measurements

### *Physical Activity Categories: Cut-point Based Summary Measures*

[2] Signatures were determined by clustering the raw signal and examining outlying groups for shared signal patterns, which were subsequently confirmed by ActiGraph’s support team as device malfunction.

Counts-per-minute (CPM) were calculated from the vector-magnitude of the GT3X triaxial count data and categorized as sedentary behavior (SB), light physical activity (LPA), and moderate to vigorous physical activity (MVPA); a combination of moderate physical activity (MPA) and vigorous physical activity (VPA). The cut-points for these categories were taken from Romanzini (Romanzini et al., 2014). Every minute from 7:00 AM to 11:00 PM was categorized in this way and the total amount of time spent in each category was tabulated per subject-day. The average amount of time spent in each physical category was derived from the tabulated activity data of each subject.

***Daily Accelerometer Summary Measures***

Accelerometer summary measures — Time Active Mean (TAM), Time Active Variability (TAV), Activity Intensity Mean (AIM), and Activity Intensity Variability (AIV) — are calculated from variation measured from the accelerometers' raw signal data as outlined in Bai et al (2014) for each day (Bai et al., 2014). Achieving this requires the use of a binary activity label, *L(t)*, which categorizes accelerometer readings at time *t* as either active (1) or inactive (0). This label is obtained on a one-second interval by comparing the standard deviation (SD) of the accelerometer signal to a cut-point, C, which is determined from the data. Bai and others did this in their data by examining the SD's density and cumulative distributions and then selecting the point at which there was a clear, visible flattening of the density. We followed the same methodology and selected C=0.6 based on our results (Supplemental **Figure S1**). TAM and TAV are, respectively, the mean and standard deviation of the time spent active (*L(t)=1*) for each person for each day examined. This gets at the duration of time spent active, but not the intensity of the activity. Activity intensity is defined as the standard deviation of the original signal relative to the average standard deviation of the signal when inactive, which is calculated per subject-day. The AIM and AIV metrics are the average and standard deviations, respectively, of the activity intensity signal during active periods. Supplementary **Table S2** and **Table S3** present these metrics for three Stanford GOALS participants to help orient readers to the dataset.

***Daily Physical Activity Patterns***

In this study, we propose a novel metric based on patterns of activity. We employed *k*-means clustering to categorize the daily activity of a participant (from 7:00 AM to 11:00 PM) based on their average counts per 10-minute intervals, calculated over consecutive, non-overlapping periods. This method generated a 96-element profile vector for each day, yielding a unique profile vector for each valid day of the week for every child (with a maximum of 7 days considered per child). **Figure 2** illustrates how these profiles can be clustered to identify common patterns of daily activity within the group. To estimate the number of clusters, *k*, we utilized the prediction strength metric developed by Tibshirani and Walther (Tibshirani & Walther, 2005), along with Yu et al.'s recommendations (Yu et al., 2023). We calculated the prediction strength statistic for up to *k*=50 and selected the highest value of *k* above 0.75 or where we observed an "elbow" (or local maxima) in the prediction strength. This approach allowed us to effectively analyze and understand pediatric activity data within the context of our clustering methodology as outlined in Yu et al. (Yu et al., 2023). In this study we use counts for comparison, but the same approach can be applied to any time series unit, such as raw data, counts, monitor independent movement summaries (MIMS), or others.

**Figure 2.** Three physical activity profiles found when clustering accelerometer count data on 10-minute intervals across the day, between 07:00 and 23:00. Cluster #1 captures participants with low activity throughout the day. Cluster #2 represents those with higher activity levels from morning until evening (17:40). Cluster #3 captures the "night-owl" group which exhibit comparably increased evening activity. The number of daily activity patterns or *shapes* that make up each cluster are listed in the plot's legend along with the number of unique participants contributing shapes to the cluster profile and the ratio of participants to the number of shapes.

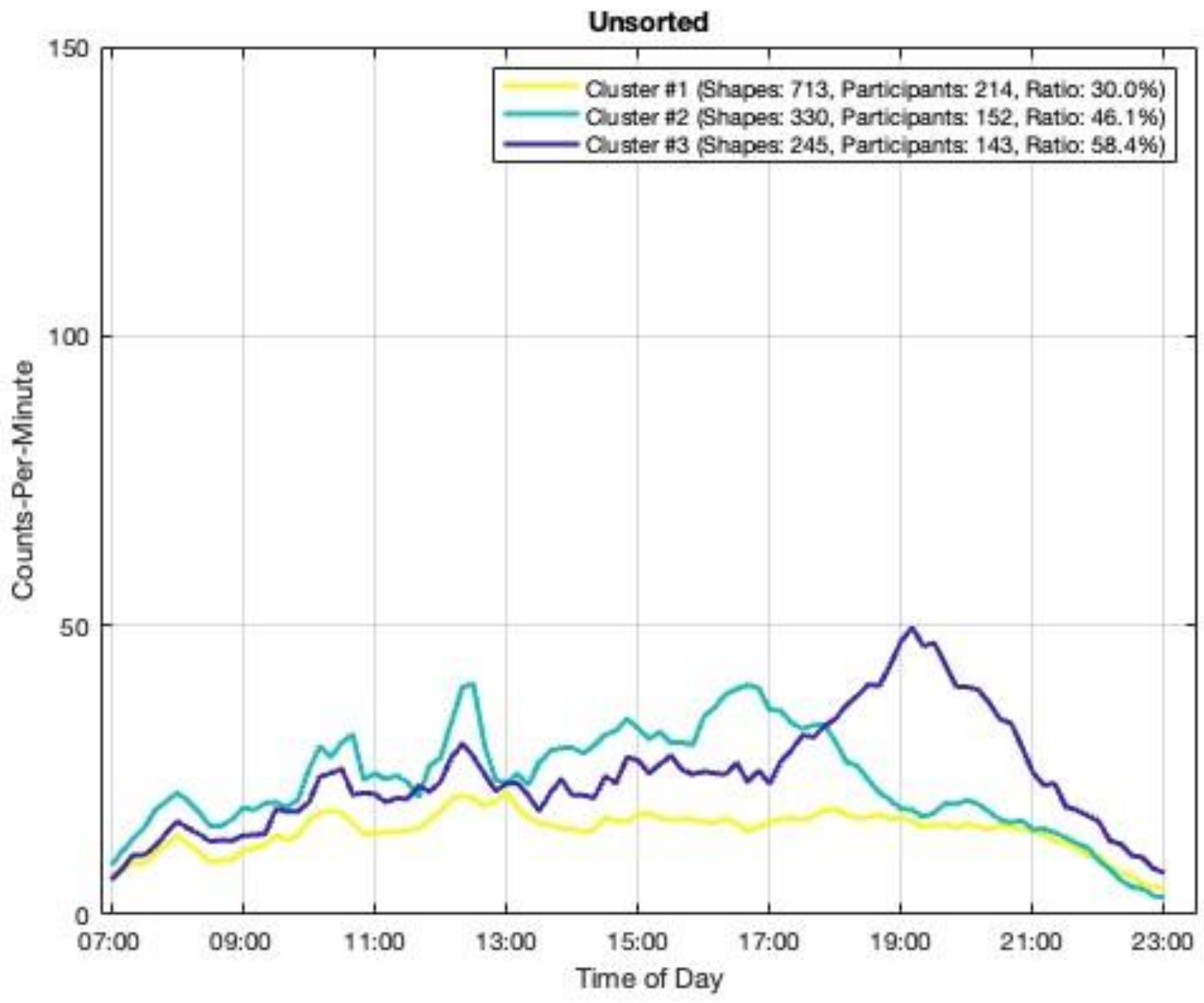

### *Sorting activity levels prior to clustering*

Minor differences in timing of activity could lead to the discovery of multiple patterns or clusters that are not meaningfully different. To address this, we considered additional ways to process the data, specifically by sorting activity levels by magnitude of intensity within pre-specified time periods while maintaining the order of those time periods (e.g., morning, afternoon, evening) prior to applying the *k*-means clustering method. For more rationale behind this thinking, consider the following. There may be a pattern reflecting high morning and low afternoon and evening activity and another reflecting low morning and evening activity coupled with high afternoon activity. Processing the data by sorting within these time periods before applying clustering methods would facilitate identifying such differences while keeping days with minor perturbations in the same cluster (e.g., peak of morning activity that lasts 10 minutes at 8:00 AM vs 8:30 AM). An extreme way to process the data – that removes the temporal features -- would be to sort the vectors from the highest to the lowest activity values *over the whole day*, allowing the discovery of clusters that distinguish days of activity by the level and length of activity irrespective of when the activity occurred. We only included this type of sorting for illustrative purposes where we sorted the day as one 16-hour long segment (Sort_01). More importantly, to maintain key temporal aspects, we considered sorting the data by intensity in a variety of ways that maintained some temporality to varying extents: by segmenting the day into two 8-hour segments (Sort_02), four 4-hour segments (Sort_04), eight 2-hour segments (Sort_08), and sixteen 1-hour segments (Sort_16) (**Figure 3**).

**Metrics for comparing summary measures**

To evaluate the candidacy of our new summary measure (and its variations by sorting approach), we compared the variation in a given outcome explained by each of the candidate summary measures using the coefficient of determination, $R^2$. We chose three key clinical outcomes: waist circumference, fasting triglyceride levels, and fasting insulin levels, to use as dependent variables in linear regression models that included accelerometer-derived physical activity measures. All models were adjusted by age and sex as a biological variable. The Akaike Information Criterion (AIC) was used to compare model performance with each of our approaches. A lower AIC is preferred.

***Modeling outcome as a function of cut-point-based summary measures***

To compare the proportion of outcome variation explained by a candidate summary measure, a linear regression framework was used. Specifically, for each summary measure, the

**Figure 3.** Daily activity profiles found by sorting accelerometer data, within consecutive segments of the day, prior to clustering. Accelerometer count data on 10-minute intervals from 07:00-23:00 were sorted for each participant prior to clustering.

**Unsorted**

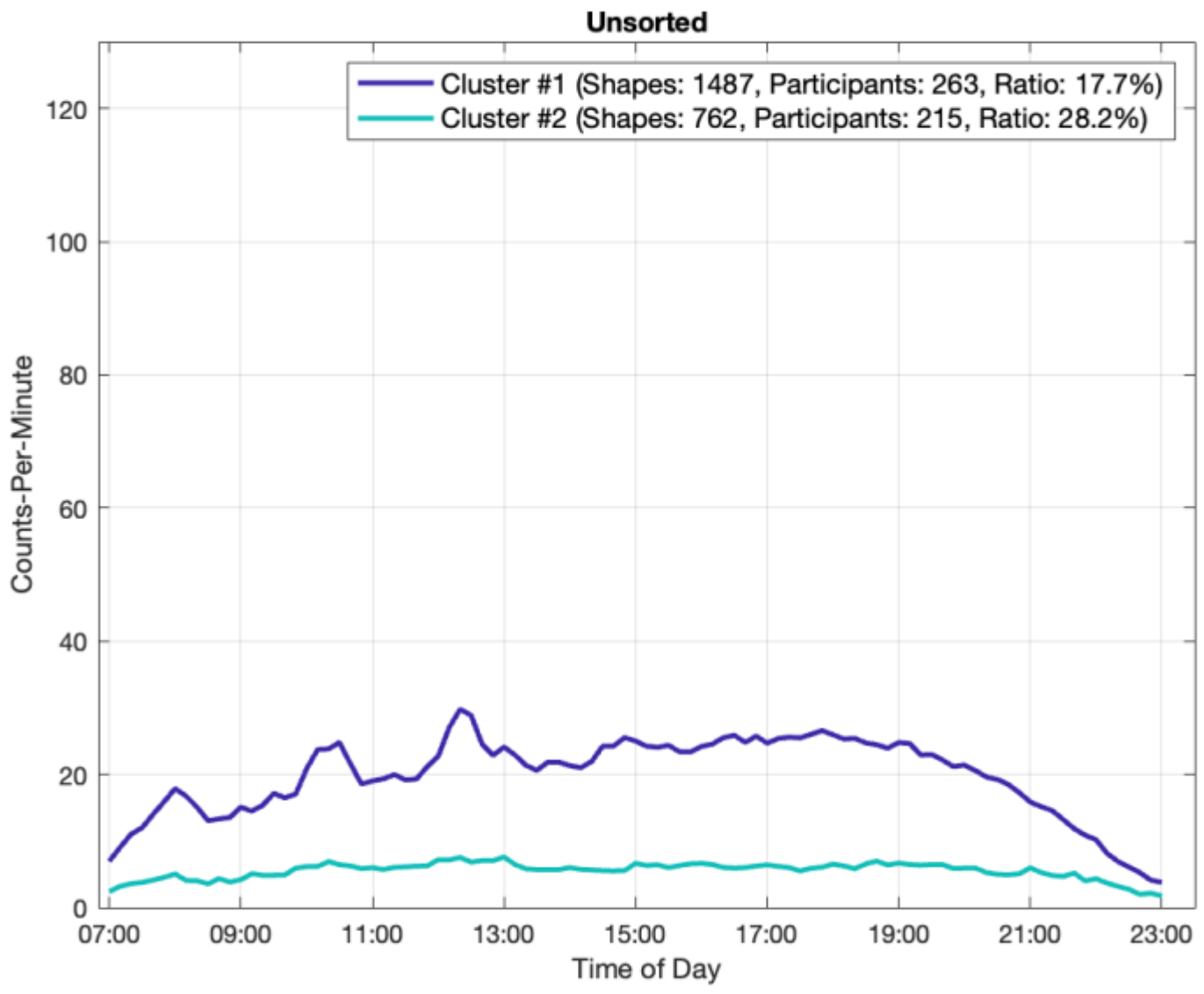

**Sort_16** Daily accelerometer data is first split into 16 consecutive 1-hour segments that are individually sorted prior to clustering.

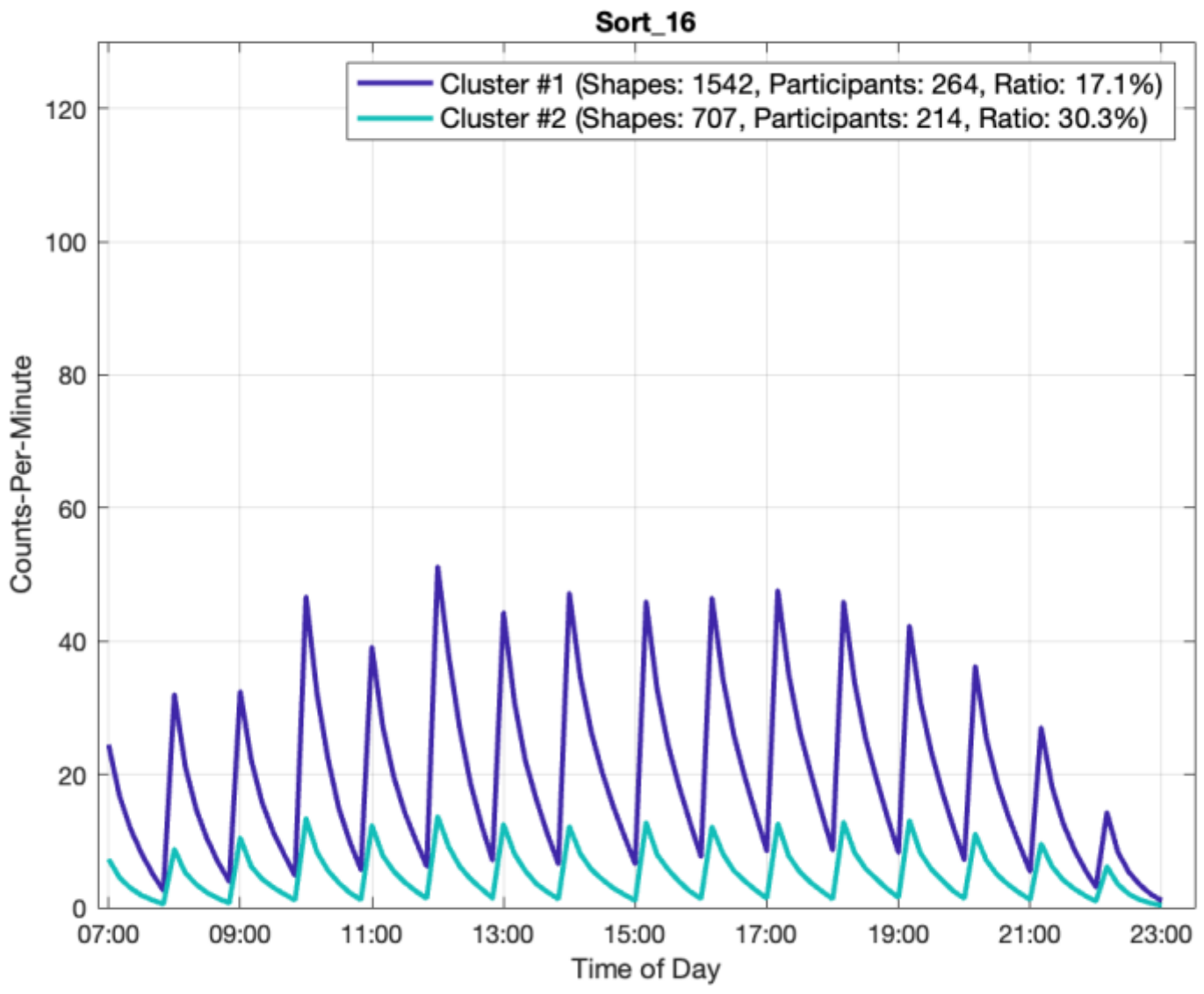

**Sort_08** Daily accelerometer data is first split into eight consecutive 2-hour segments that are each sorted.

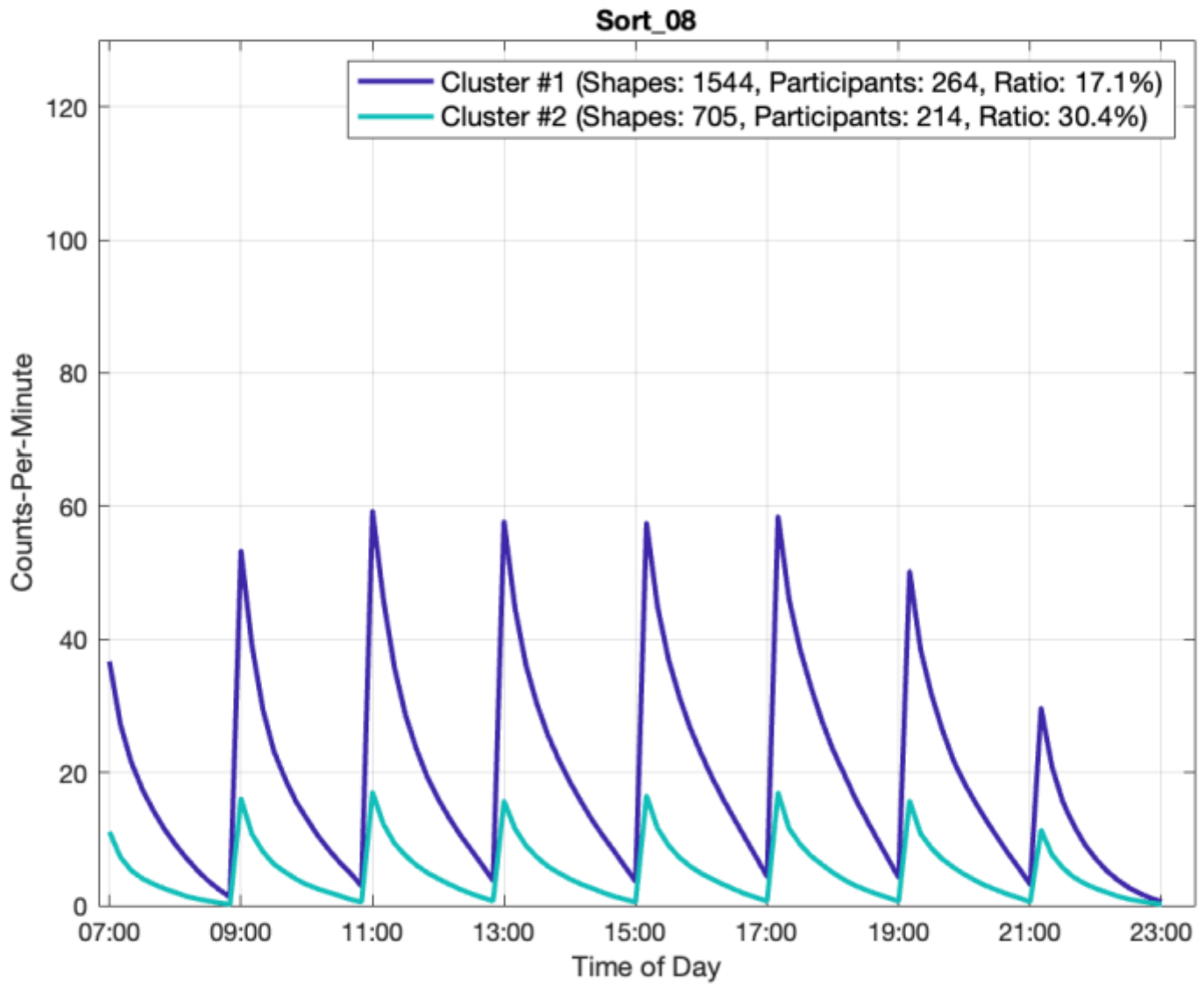

**Sort_04** Daily accelerometer data is split into four consecutive 4-hour segments which are locally sorted prior to clustering.

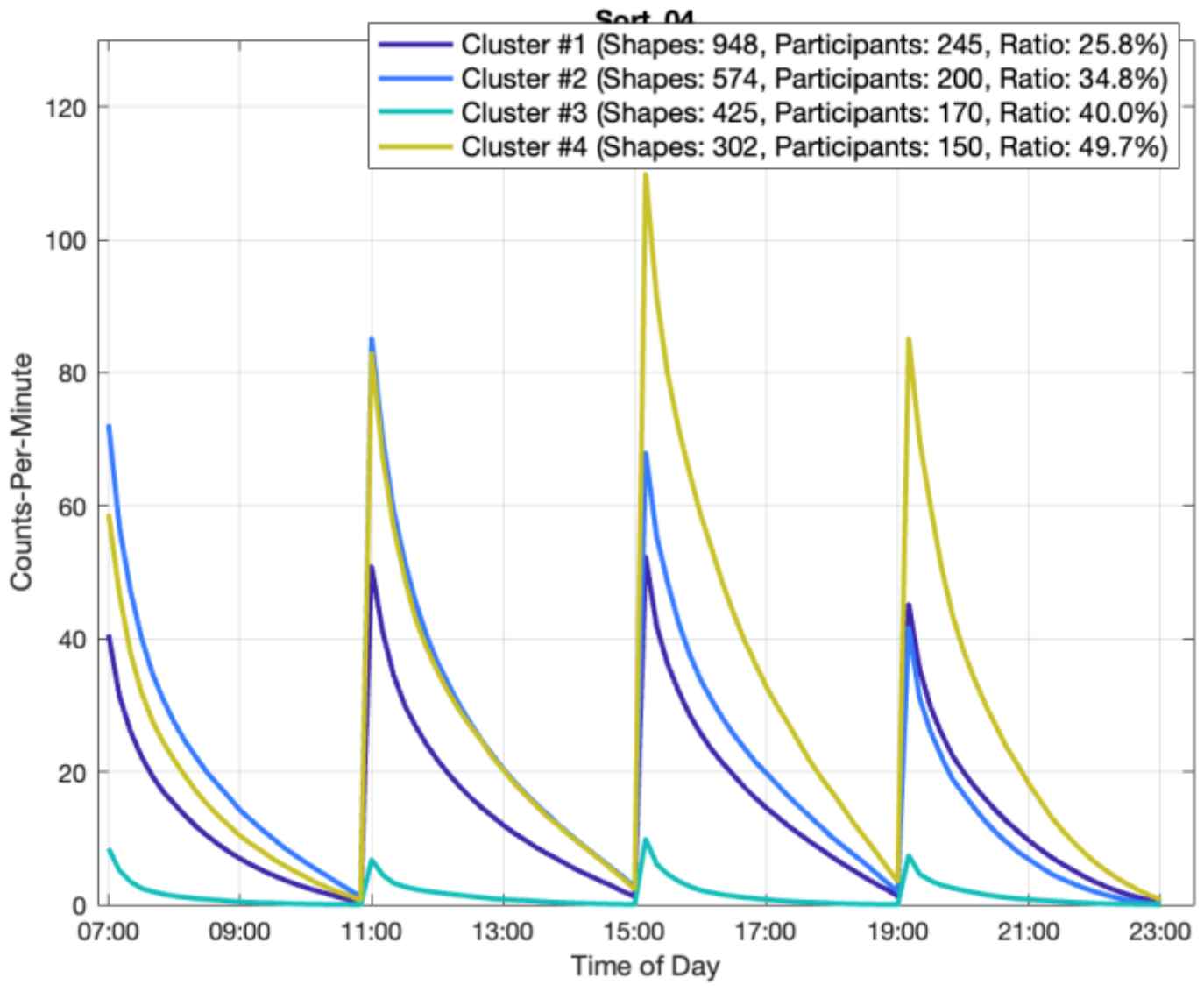

| | |
|---|---|
| **Sort_02** Daily accelerometer data is split into two consecutive 8-hour segments which are each sorted prior to clustering. 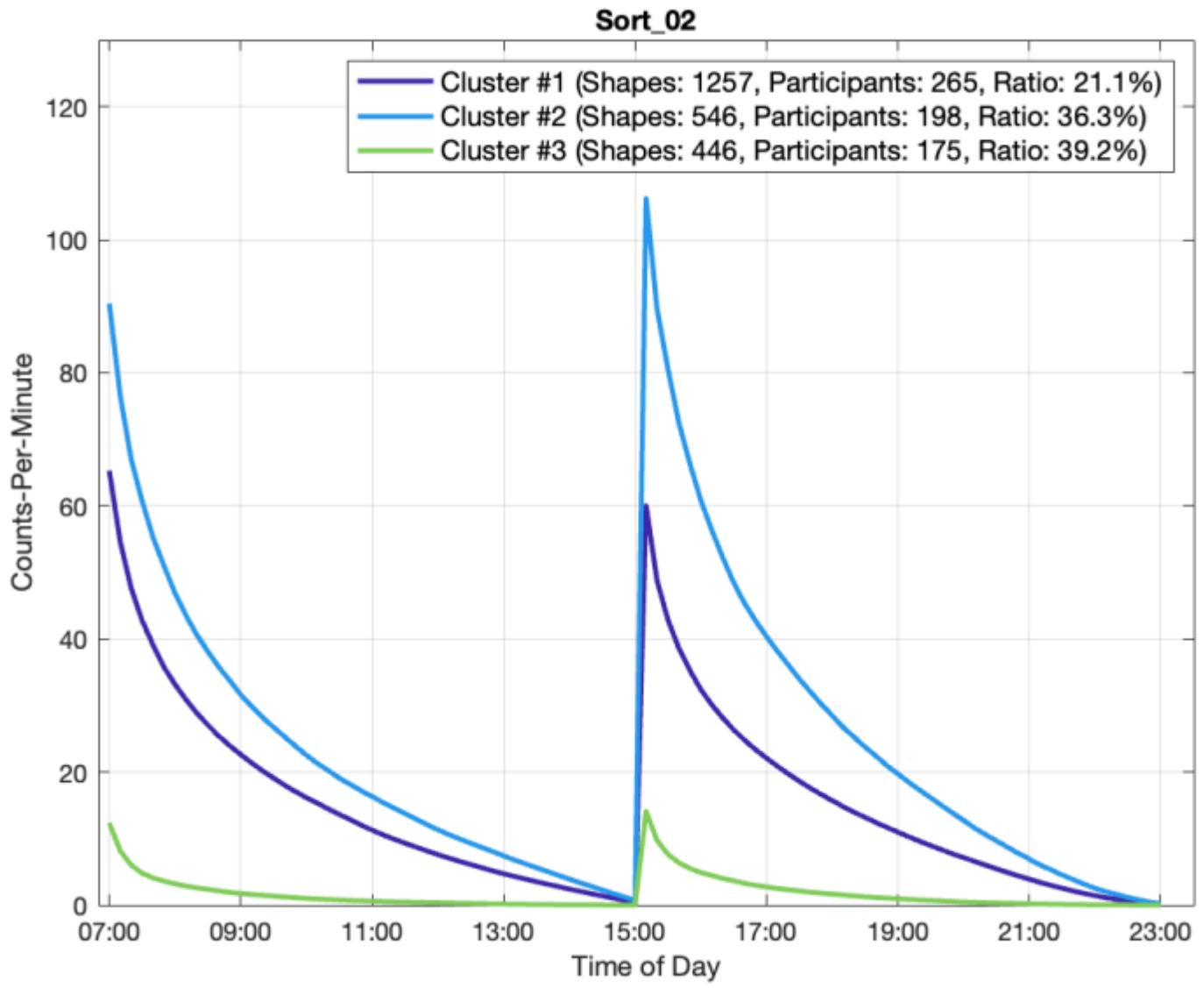 | **Sort_01** Accelerometer data is sorted across the entire day prior to clustering. Temporal information is discarded in this scenario. 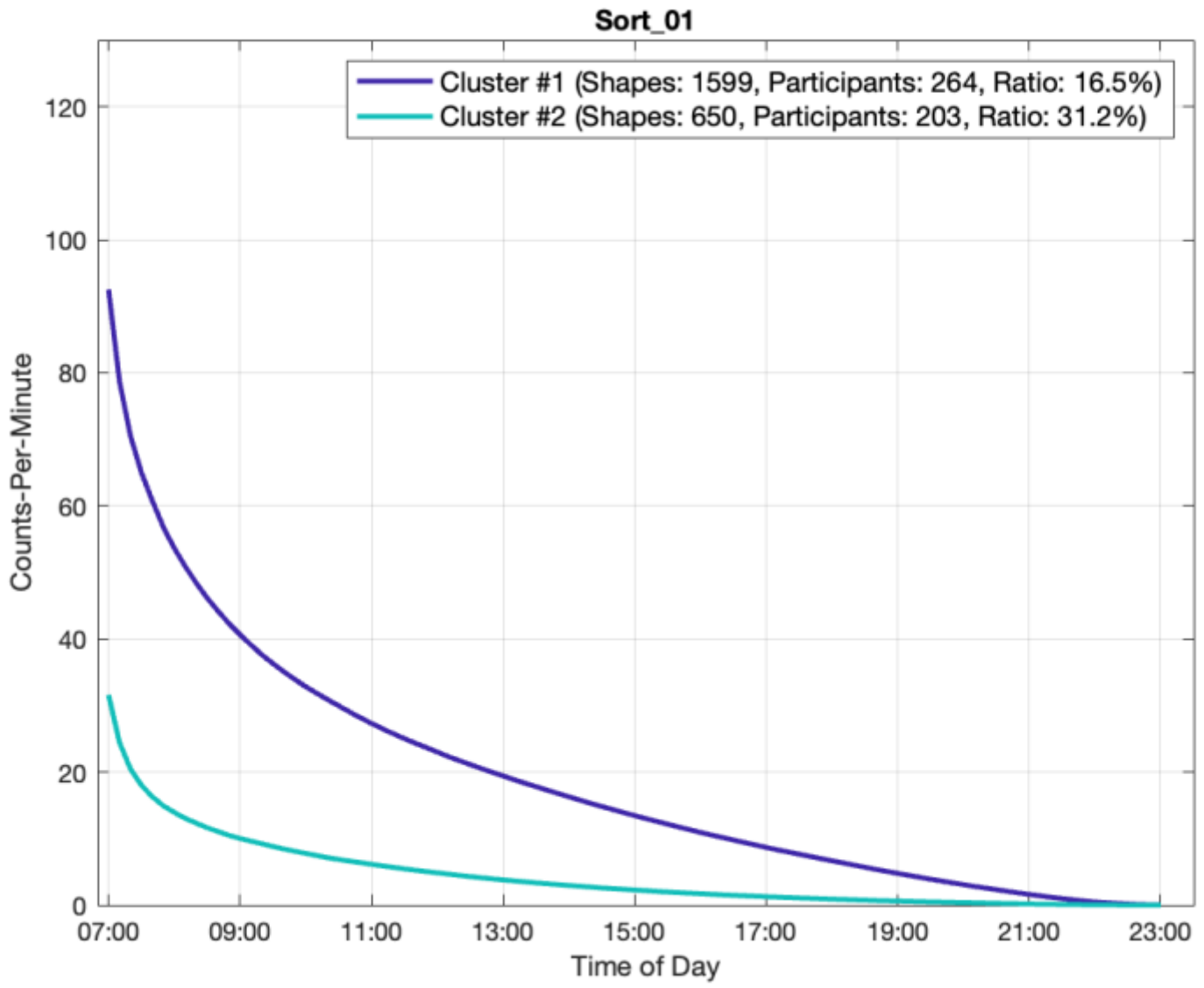 |

outcome as a function of the proportion of observed days assigned to a particular cluster as shown in equation 1

$$Y_i = \beta_0 + \beta_1 \cdot A_i + \beta_2 \cdot S_i + \sum_{j=1}^{k} \beta_{j+2} \cdot C_{j,i} \; + \; \varepsilon_i \qquad (1)$$

where $C_{j,i}$ is the proportion of days that participant *i* had in cluster *j* relative to the total number of days observed for participant *i*, such that $\sum_{j=1}^{k} C_{j,i} = 1$. For example, if the total clusters are *k*=3 and a given participant had seven days of activity, with four days belonging to cluster #1, three days belonging to cluster #2, and zero days belonging to cluster #3, then their cluster membership distribution would be $C_1$=4/7=0.5714, $C_2$=3/7=0.4286, and $C_3$=0. Thus, we are not constrained by the total number of clusters or differences in the total number of days available for each participant, which may vary (e.g., removal due to data cleaning), because they are normalized according to the participant's total contributions (seven in the example given). This modeling approach was applied to the clusters defined by each of our preprocessing approaches: unsorted and sorted, in day-wise partitions of 1, 2, 4, 8, and 16 hours.

## RESULTS

**Table 2** presents variance explained ($R^2$) and goodness of fit (AIC) corresponding to each candidate summary measure when modeling waist circumference, fasting insulin levels, and fasting triglyceride levels in Stanford GOALS participants; $R^2$ and AIC values for a model using only age and sex are included for reference as well. Results were comparable across measures within each outcome considered. For example, the proportion of variation in insulin explained ranged from only 0.0636 (LPA) to 0.1064 (VPA) for the physical activity categories. The daily accelerometer summary measures have a similar level of variation explained and range from 0.0618 (AIV) to 0.1024 (TAM). Cluster member distribution explains from 0.0872 (unsorted clusters) to 0.1011 (sorted in 4-hour intervals). Further, the ability for all measures to explain variation varied widely depending on the clinical outcome. For example, the unsorted cluster membership distribution explains only 0.0221 of the variation observed in triglycerides, 0.0872 of insulin variation, and a proportion of 0.2300 in waist circumference variation. For all measures, the variation in waist circumference was explained best (highest $R^2$ values and lowest AIC) using each approach, followed by fasting insulin and then fasting triglyceride levels.

We additionally evaluated how alternative pre-processing or sorting methods prior to clustering influenced model performance for each outcome. When modeling with cluster membership distribution, variation explained was comparable across sorting approaches. However, sorting the days in eight 2-hour sections (Sort_08) prior to clustering yielded the best results for waist circumference ($R^2$=0.2516, AIC=1880.5) while presorting in two 8-hour sections (Sort_02) gave the best results for insulin ($R^2$=0.1051, AIC=1949.4) and presorting within four 4-hour sections was best for triglycerides ($R^2$=0.0585, AIC=2763.6).

When modeling with physical activity categories, duration of SB was best for waist circumference ($R^2$ =0.2422, AIC=1881.7) while duration of VPA was best for insulin ($R^2$ =0.1064, AIC=1945.0) and triglycerides ($R^2$ =0.0288, AIC=2765.5). When using Bai's physical summary measures, TAM performed best for modeling each outcome: waist circumference ($R^2$=0.2539, AIC=1850.8), insulin ($R^2$=0.1024 and AIC=1913.8), and triglycerides ($R^2$=0.0555, AIC=2733.5).

## DISCUSSION

Summarizing accelerometer-derived physical activity patterns in a meaningful way remains a challenge in the field. Traditional summary measures, including cut-point-based classifications and intensity-derived metrics (Bai et al., 2014), capture key aspects of  physical activity but may overlook important temporal patterns that influence health outcomes. Recent studies leveraging clustering techniques have demonstrated that *when* physical activity occurs can be just as relevant as how much activity is performed (Aqeel et al., 2021; Nawrin et al., 2024a; Niemelä et al., 2019; Smagula et al., 2022). Our findings contribute to this growing body of research by demonstrating that cluster-based summary measures can explain a comparable proportion of variation in metabolic outcomes while preserving key temporal aspects of physical activity behavior.

**Table 2.** Mean time spent within physical activity categories, accelerometry summary measures, and cluster membership as predictors of waist circumference, fasting insulin, and fasting triglycerides in Stanford GOALS participants covarying for age and sex. Activity categories include sedentary behavior (SB), light physical activity (LPA), moderate to vigorous physical activity (MVPA), moderate physical activity (MPA), and vigorous physical activity (VPA). Accelerometer summary measures include time active mean (TAM), time active variability (TAV), activity intensity mean (AIM), and activity intensity variability (AIV). Clustering was performed on consecutive mean count values derived from 10-minute, non-overlapping intervals from 7:00AM to 11:00PM each day. See the text for explanation of the different sorting approaches applied prior to clustering.

| | | Age+Sex Covariate | Physical Activity Categories | | | | | Accelerometer Summary Measures | | | | Cluster Membership | | | | | |
|---|---|---|---|---|---|---|---|---|---|---|---|---|---|---|---|---|---|
| Health Outcome | Statistic | | SB | LPA | MVPA | MPA | VPA | TAM | TAV | AIM | AIV | Unsorted | Sort_01 | Sort_02 | Sort_04 | Sort_08 | Sort_16 |
| Waist (cm) | $R^2$ | 0.1911 | 0.2422 | 0.2332 | 0.2183 | 0.2039 | 0.2238 | 0.2539 | 0.2418 | 0.2349 | 0.2149 | 0.2300 | 0.2348 | 0.2281 | 0.2190 | 0.2516 | 0.2266 |
| | AIC | 1869.2 | 1881.7 | 1884.7 | 1889.6 | 1894.3 | 1887.8 | 1850.8 | 1854.9 | 1857.2 | 1863.7 | 1887.8 | 1886.1 | 1890.4 | 1895.4 | 1880.5 | 1888.9 |
| Insulin | $R^2$ | 0.0522 | 0.0786 | 0.0636 | 0.0953 | 0.0735 | 0.1064 | 0.1024 | 0.0956 | 0.0796 | 0.0618 | 0.0872 | 0.0970 | 0.1051 | 0.1011 | 0.0995 | 0.0937 |
| | AIC | 1925.5 | 1952.8 | 1957.0 | 1948.1 | 1954.2 | 1945.0 | 1913.8 | 1915.7 | 1920.1 | 1925.0 | 1952.4 | 1949.7 | 1949.4 | 1952.5 | 1949.0 | 1950.6 |
| Triglyceride | $R^2$ | 0.0143 | 0.0238 | 0.0223 | 0.0256 | 0.0225 | 0.0288 | 0.0555 | 0.0484 | 0.0449 | 0.0300 | 0.0221 | 0.0227 | 0.0344 | 0.0585 | 0.0230 | 0.0221 |
| | AIC | 2742.1 | 2766.9 | 2767.3 | 2766.4 | 2767.2 | 2765.5 | 2733.5 | 2375.2 | 2376.2 | 2740.1 | 2769.3 | 2769.2 | 2768.1 | 2763.6 | 2769.1 | 2769.3 |

Our study demonstrated comparable findings in proportion of variation explained by a cluster-based physical activity measure vs by traditional summary metrics of physical activity, suggesting the promise of unsupervised clustering techniques in deriving such measures for the purpose of studying the role of physical activity in health. Although the absolute proportion of variance explained is modest and provided primarily for illustrative purposes, the key point of interest is the comparative performance. Importantly, the choice of summary metric should be driven primarily through the research question.

Sorting the data by the signal's magnitude prior to clustering had implications in terms of interpretation. Without sorting, the regression parameter corresponding to a given cluster indicates that an increase in days spent with a given activity pattern is associated with an increase or decrease in the health outcome. With sorting, the interpretation takes on a different meaning. For example, if the entire day is sorted, it means increases in days that have attained a certain level of sustained activity are associated with increases/decreases in the health measure, with a loss in the temporal aspect of the measure. Such a measure becomes much more like that of Bai's daily accelerometer measures (See **Supplementary Table S4** for a variance-covariance matrix of all summary measures considered here). Sorting within time segments, however, allows questions about intensity of activity at specific periods within a day. For example, one cluster may describe days that have brief high activity levels in the morning and sustained flat shapes in other periods, whereas another cluster may have similar shapes for all periods in the day with no distinctive shape for the morning period. The parameters corresponding to these clusters will indicate the differences in strength and magnitude of associations between such behaviors. These findings highlight the importance of tailoring modeling approaches to the specific questions being addressed (e.g., do we hypothesize that it is a certain amount of intense activity whenever it occurs or are we interested in seeing differences by morning, afternoon, evening?).

Our work stems from the promise around high volumes of data generated continuously from accelerometers that allow us to answer questions limited by traditional measures. While traditional methods capture important dimensions of physical activity, they do not reflect the sequence and timing of movement across the day. A cluster-based approach complements these measures by allowing researchers to model not just intensity and variability, but also temporal consistency and shifts in physical activity behaviors. Our work is motivated by other studies that demonstrate keen interest in understanding the role of the timing and order of physical activity. Studies have consistently found that distinct morning, evening, and all-day physical activity profiles are linked to different health risks (Aqeel et al., 2021; Nawrin et al., 2024b; Niemelä et al., 2019). Our results similarly suggest that representing physical activity as a distribution of cluster membership can explain variation in key clinical outcomes suggesting additional insights beyond total physical activity volume. Additionally, research has shown that later activity patterns and irregular physical activity rhythms are associated with higher depression risk and poorer metabolic profiles (Nawrin et al., 2024b; Smagula et al., 2022). Our study reinforces this by highlighting how different sorting strategies prior to clustering can help refine questions to investigate associations with sustained activity levels versus temporally localized bursts of activity. Importantly, the differences in findings by different sorting approaches demonstrates that not only are there implications for the shaping of the research question but also for replicability of findings across studies. This raises a critical issue that accompanies a cluster-derived approach – the various choices the researcher makes when deriving the summary metric which could have downstream effects on the findings. Jones and others (2021) discuss this in their systematic review and emphasize the importance of standardizing choices from the underlying machine learning method to how parameters are tuned.  We agree with these remarks and additionally raise the importance of how missing data are handled and how the data are processed. While we did not illustrate different approaches for handling missing data in this paper, we focused on some pre-processing approaches in the form of sorting methods. We urge investigators to pre-specify their derivation plan including pre-processing and missing data in their statistical analysis plan. Doing so will increase rigor, ensure reproducibility, and facilitate replicability.

A major strength of our study is the examination and application of sorting-based pre-processing prior to clustering, which refines pattern recognition by distinguishing sustained versus transient activity states. Additionally, the regression framework offers a formal method for linking temporal patterns to health outcomes. Our results suggest that sorting activity levels into structured time segments before clustering enhances interpretability and model fit, reinforcing the importance of data pre-processing decisions in machine learning-based physical activity research.

Our study is not without limitations. First, we demonstrate findings on a limited data set of one week of accelerometer signal per participant. Having multiple weeks of data might reveal greater heterogeneity in performance across approaches. Second, we constrained our days' timeframe to minimize the amount of data excluded due to non-wear or device malfunction. The performance of these methods may depend on the method used to handle missing data. Our team is currently exploring how to adapt methods for multiple imputation when deriving cluster-based predictors in a regression framework. While we evaluated one processing characteristic of the clustering approach (i.e., sorting the data), there are other processing characteristics that may influence performance of this approach. For instance, an essential element of any unsupervised clustering method is the a priori selection of the number of clusters, which fundamentally shapes the analytical outcomes. Here we relied on the prediction strength metric to determine the optimal number of clusters; this approach requires repeatedly splitting the input data into testing and evaluation sets to measure how well data fall into the clusters for each value of $k$ considered. However, the number of clusters could be determined by other methods, such as the Calinski-Harabasz index or the silhouette index (Caliński & Harabasz, 1974; Rousseeuw, 1987). While we demonstrated principles using count data, an appealing feature of the clustering approach is that it can be readily applied to accelerometer data on other scales. This includes, for example, the monitor-independent movement summary (MIMS) scale, which is an open-source alternative to counts, as well as the raw acceleration (gravity) signal. An important limitation of the newly proposed summary metric is that it may be sample dependent. Patterns that are discovered may be unique to the data set and sensitive to variations in that dataset. Thus, when performing hypothesis testing on relationships between patterns and outcome, one may wish to leverage re-sampling or bootstrapping approaches that can capture the uncertainty of the discovered clusters. This remains future work.

New accelerometer summary measures may enable addressing many more types of questions about the role of physical activity. Further, with more real-world data being generated per participant, we may broaden how we address questions. Specifically, we may consider discovering relationships about the role of physical activity that are specific to the individual. Suppose that the role of physical activity affects one individual's mood differently than another. We may then consider a different set of cluster-derived predictors for each individual to address whether and how the role of physical activity affects mood. In such cases, it may make more sense to identify patterns of activity within an individual to enable addressing questions in a manner that is highly tailored to an individual (e.g., are changes in pattern relative to the individual's typical pattern associated with changes in a key outcome).

While our primary interest was in physical activity patterns based on accelerometer measurements, this approach is not limited to this context alone. The method we propose can be applied to profiling other patterns of human behavior derived from various sensors, such as smartwatches that monitor heart rate variability or electroencephalogram (EEG) devices that track brain activity. This broad applicability underscores the versatility and potential of our summary metric when relating patterns to outcome in various fields of medical and behavioral research.

## SUMMARY

We build on the work of other researchers who have leveraged unsupervised clustering to examine physical activity patterns and to further relate them to health outcomes. Our study demonstrated the promise of such techniques in deriving meaningful measures of physical activity patterns that can then be leveraged to understand the role of physical activity in health and clinical outcomes. Still, there are important challenges to acknowledge that remain key topics for future research. These challenges – how to handle missing data, how to pre-process the data, how to standardize the machine learning approach – pose threats to the replicability of studies. For this purpose, we urge researchers to pre-specify their choices in deriving the summary metric and to report them with transparency to facilitate rigor, reproducibility, and replicability. Finally, although we illustrated ideas using accelerometer data, the principles discussed can be applied to deriving other types of patterns based on intensively sampled time series data.

**Acknowledgements and Current Funding Sources/Financial Disclosure**

Research reported in this publication was supported in part by grants from the Stanford Maternal and Child Health Research Institute, the Department of Pediatrics at Stanford University, and the Li Ka Shing Foundation Stanford–Oxford Big Data for Human Health seed grant program. The content is solely the responsibility of the authors and does not necessarily represent the official views of Stanford University or the Li Ka Shing Foundation.

This manuscript is also partially supported by the following NIH grants: R01LM013355, *Novel machine learning and missing data methods for improving estimates of physical activity, sedentary behavior, and sleep using accelerometer data*; UL1TR003142, Stanford's Center for Clinical and Translational Education and Research (CTSA) under the Biostatistics, Epidemiology, and Research Design (BERD) Program; P30DK116074, the Clinical and Translational Core of the Stanford Diabetes Research Center; and P30CA124435, the Biostatistics Shared Resource (B-SR) of the NCI-sponsored Stanford Cancer Institute.

Data were collected with funding from the National Heart, Lung, and Blood Institute of the National Institutes of Health under award number U01HL103629.

The content expressed in this paper is solely the responsibility of the authors and does not necessarily represent the official views of the National Heart, Lung, and Blood Institute; the National Institutes of Health; the U.S. Department of Health and Human Services; or the U.S. Government.

We thank the children and families who participated.

**SUPPLEMENTARY MATERIAL**

**Figure S1.** The density curves of standard deviations for 252 subjects covering 7 days each. The upper panel contains density curves of the original standard deviations and the lower panel contains density curves of standard deviations greater than c=0.6. The vertical black lines in both panels are at x=0.6. (Verbiage taken from Bai, 2014, with our results written in.)

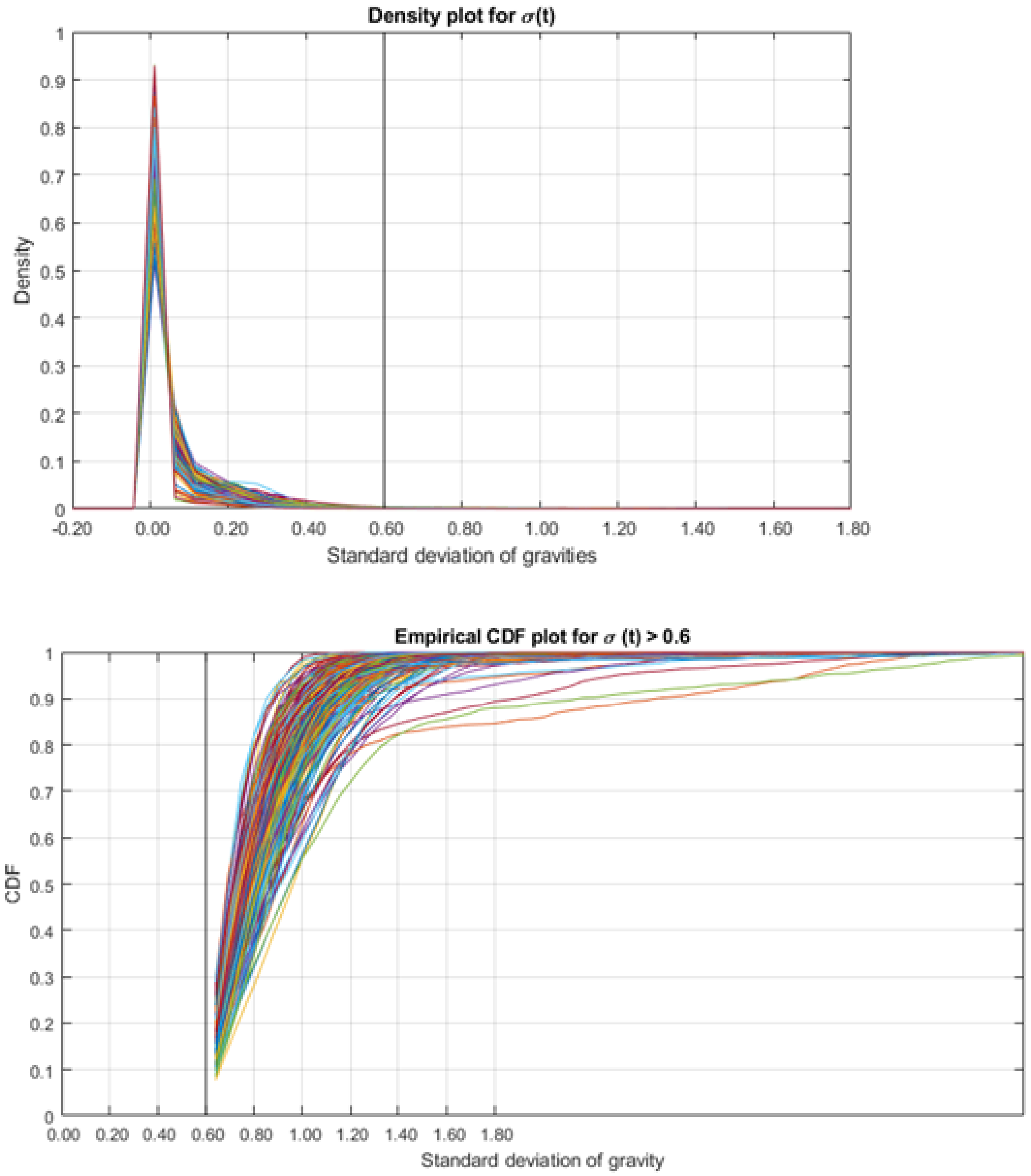

**Table S1**: Time of day selection for data quality. Choi's method was employed to identify non-wear at 1-minute intervals in the count data (Choi, 2011). We used an iterative clustering scheme on the raw signal data (i.e. gravity) and examined the profiles of outlying groups for signatures of device malfunction. These signatures were confirmed via email correspondence with ActiGraph's technical support team. Confirmed malfunctions included spikes of activity outside the normal range and sensors producing the same non-zero value for prolonged periods, indicative of sensor sticking.

| Time frame | Subjects with 0 profiles (n) | Subjects with 0+ profiles (n) | Subjects with 1+ profiles (n) | Subjects with 2+ profiles (n) | Subjects with 3+ profiles (n) | Subjects with 4+ profiles (n) | Subjects with 5+ profiles (n) | Subjects with 6+ profiles (n) |
|---|---|---|---|---|---|---|---|---|
| 05:00-17:00 | 10 | 258 | 255 | 252 | 246 | 226 | 203 | 135 |
| 05:00-18:00 | 10 | 258 | 255 | 252 | 240 | 223 | 202 | 129 |
| 05:00-19:00 | 10 | 258 | 255 | 250 | 237 | 220 | 197 | 120 |
| 05:00-20:00 | 10 | 258 | 254 | 247 | 236 | 217 | 191 | 105 |
| 05:00-21:00 | 11 | 257 | 253 | 246 | 229 | 213 | 179 | 82 |
| 05:00-22:00 | 12 | 256 | 251 | 240 | 222 | 199 | 151 | 60 |
| 05:00-23:00 | 13 | 255 | 245 | 232 | 211 | 177 | 117 | 38 |
| 05:00-24:00 | 15 | 253 | 244 | 227 | 201 | 163 | 92 | 29 |
| 06:00-18:00 | 9 | 259 | 256 | 252 | 242 | 224 | 205 | 133 |
| 06:00-19:00 | 9 | 259 | 256 | 250 | 239 | 221 | 200 | 124 |
| 06:00-20:00 | 10 | 258 | 255 | 247 | 238 | 218 | 194 | 109 |
| 06:00-21:00 | 11 | 257 | 254 | 246 | 231 | 214 | 182 | 86 |
| 06:00-22:00 | 11 | 257 | 251 | 240 | 225 | 200 | 154 | 62 |
| 06:00-23:00 | 13 | 255 | 245 | 233 | 214 | 178 | 119 | 38 |
| 06:00-24:00 | 15 | 253 | 244 | 228 | 203 | 164 | 94 | 30 |
| 07:00-19:00 | 7 | 261 | 257 | 253 | 239 | 222 | 206 | 129 |
| 07:00-20:00 | 7 | 261 | 255 | 249 | 238 | 220 | 199 | 114 |
| 07:00-21:00 | 9 | 259 | 254 | 247 | 232 | 216 | 188 | 88 |
| **07:00-22:00** | **9** | **259** | **251** | **241** | **228** | **203** | **159** | **62** |
| **07:00-23:00** | **12** | **256** | **245** | **234** | **218** | **182** | **121** | **38** |
| 07:00-24:00 | 15 | 253 | 244 | 229 | 208 | 165 | 96 | 30 |
| 08:00-20:00 | 5 | 263 | 261 | 255 | 247 | 229 | 207 | 118 |
| 08:00-21:00 | 5 | 263 | 259 | 251 | 244 | 221 | 192 | 93 |
| 08:00-22:00 | 6 | 262 | 256 | 247 | 235 | 207 | 164 | 64 |
| 08:00-23:00 | 7 | 261 | 249 | 238 | 224 | 185 | 126 | 38 |
| 08:00-24:00 | 10 | 258 | 245 | 235 | 212 | 168 | 100 | 30 |
| 09:00-21:00 | 3 | 265 | 260 | 255 | 247 | 223 | 195 | 94 |
| 09:00-22:00 | 4 | 264 | 257 | 252 | 237 | 209 | 165 | 64 |
| 09:00-23:00 | 7 | 261 | 252 | 241 | 227 | 186 | 126 | 39 |
| 09:00-24:00 | 10 | 258 | 248 | 238 | 213 | 169 | 100 | 31 |
| 10:00-22:00 | 3 | 265 | 257 | 253 | 238 | 210 | 167 | 65 |
| 10:00-23:00 | 6 | 262 | 252 | 241 | 228 | 188 | 127 | 40 |
| 10:00-24:00 | 9 | 259 | 248 | 238 | 213 | 170 | 101 | 32 |
| 11:00-23:00 | 6 | 262 | 253 | 242 | 229 | 191 | 128 | 42 |
| 11:00-24:00 | 8 | 260 | 249 | 238 | 213 | 173 | 102 | 34 |
| 12:00-24:00 | 6 | 262 | 250 | 239 | 213 | 173 | 104 | 35 |

**Table S2:** Health outcomes and daily activity summary measures of a selection of Stanford GOALS participants.

| Participant | Age (years) | Is Male? | Day of Week | Day | Waist (cm) | Triglyceride (mg/dL) | Insulin (μIU/mL) | TAM | TAV | AIM | AIV | SB (min) | LPA (min) | MPA (min) | VPA (min) | MVPA (min) |
|---|---|---|---|---|---|---|---|---|---|---|---|---|---|---|---|---|
| 1 | 10.46 | 0 | 0 | Sun | 76.40 | 95 | 9.6 | 0.00270 | 0.05193 | 0.00042 | 0.01458 | 567 | 315 | 66 | 12 | 78 |
| 1 | 10.46 | 0 | 1 | Mon | 76.40 | 95 | 9.6 | 0.00220 | 0.04689 | 0.00022 | 0.00959 | 694 | 202 | 50 | 14 | 64 |
| 1 | 10.46 | 0 | 3 | Wed | 76.40 | 95 | 9.6 | 0.00100 | 0.03161 | 0.00003 | 0.00281 | 698 | 213 | 45 | 3 | 48 |
| 1 | 10.46 | 0 | 5 | Fri | 76.40 | 95 | 9.6 | 0.00361 | 0.05998 | 0.00031 | 0.01046 | 608 | 267 | 59 | 26 | 85 |
| 1 | 10.46 | 0 | 6 | Sat | 76.40 | 95 | 9.6 | 0.00093 | 0.03042 | 0.00006 | 0.00436 | 747 | 188 | 21 | 4 | 25 |
| 2 | 10.16 | 1 | 0 | Sun | 102.00 | 124 | 25.4 | 0.00865 | 0.09259 | 0.00088 | 0.01731 | 467 | 300 | 114 | 79 | 193 |
| 2 | 10.16 | 1 | 1 | Mon | 102.00 | 124 | 25.4 | 0.00291 | 0.05384 | 0.00019 | 0.00822 | 525 | 305 | 95 | 35 | 130 |
| 2 | 10.16 | 1 | 2 | Tue | 102.00 | 124 | 25.4 | 0.00406 | 0.06356 | 0.00030 | 0.01088 | 540 | 285 | 93 | 41 | 134 |
| 2 | 10.16 | 1 | 3 | Wed | 102.00 | 124 | 25.4 | 0.00291 | 0.05384 | 0.00025 | 0.01151 | 624 | 222 | 78 | 36 | 114 |
| 2 | 10.16 | 1 | 5 | Fri | 102.00 | 124 | 25.4 | 0.00931 | 0.09606 | 0.00046 | 0.01111 | 478 | 304 | 107 | 71 | 178 |
| 2 | 10.16 | 1 | 6 | Sat | 102.00 | 124 | 25.4 | 0.01320 | 0.11415 | 0.00221 | 0.03207 | 425 | 290 | 118 | 127 | 245 |
| 3 | 11.03 | 0 | 0 | Sun | 119.30 | 73 | 28.2 | 0.00217 | 0.04650 | 0.00025 | 0.01011 | 506 | 363 | 73 | 17 | 90 |
| 3 | 11.03 | 0 | 1 | Mon | 119.30 | 73 | 28.2 | 0.00200 | 0.04468 | 0.00005 | 0.00258 | 590 | 295 | 37 | 38 | 75 |
| 3 | 11.03 | 0 | 3 | Wed | 119.30 | 73 | 28.2 | 0.00120 | 0.03467 | 0.00011 | 0.00536 | 526 | 349 | 66 | 17 | 83 |
| 3 | 11.03 | 0 | 4 | Thur | 119.30 | 73 | 28.2 | 0.00104 | 0.03219 | 0.00008 | 0.00504 | 622 | 263 | 56 | 18 | 74 |
| 3 | 11.03 | 0 | 5 | Fri | 119.30 | 73 | 28.2 | 0.00354 | 0.05937 | 0.00038 | 0.01098 | 561 | 277 | 92 | 30 | 122 |
| 3 | 11.03 | 0 | 6 | Sat | 119.30 | 73 | 28.2 | 0.00078 | 0.02788 | 0.00000 | 0.00028 | 554 | 334 | 65 | 7 | 72 |
| : | : | : | : | | : | : | : | : | : | : | : | : | : | : | : | : |

**Table S3:** Health outcomes and average of daily activity summary measures of a selection of Stanford GOALS participants.

| Participant | Age (years) | Is Male? | Waist (cm) | Triglyceride (mg/dL) | Insulin (μIU/mL) | Mean TAM (min) | Mean TAV (min) | Mean AIM (min) | Mean AIV (min) | Mean SB (min) | Mean LPA (min) | Mean MPA (min) | Mean VPA (min) | Mean MVPA (min) |
|---|---|---|---|---|---|---|---|---|---|---|---|---|---|---|
| 1 | 10.46 | 0 | 76.40 | 95 | 9.6 | 0.00209 | 0.04417 | 0.00021 | 0.00836 | 662.80 | 237.00 | 48.20 | 11.80 | 60.00 |
| 2 | 10.16 | 1 | 102.00 | 124 | 25.4 | 0.00684 | 0.07901 | 0.00072 | 0.01518 | 509.83 | 284.33 | 100.83 | 64.83 | 165.67 |
| 3 | 11.03 | 0 | 119.30 | 73 | 28.2 | 0.00179 | 0.04088 | 0.00015 | 0.00573 | 559.83 | 313.50 | 64.83 | 21.17 | 86.00 |
| : | : | : | : | : | : | : | : | : | : | : | : | : | : | : |

**Table S4**: Pearson cross-correlation of summary measures.

| | Waist | Triglyceride | Insulin | TAM | TAV | AIM | AIV | SB | LPA | MPA | VPA | MVPA |
|---|---|---|---|---|---|---|---|---|---|---|---|---|
| **Waist** | 1.0000 | 0.1965 | 0.4851 | -0.3362 | -0.3353 | -0.2525 | -0.2182 | 0.3839 | -0.3361 | -0.2407 | -0.2617 | -0.2758 |
| **Triglyceride** | 0.1965 | 1.0000 | 0.3403 | -0.1561 | -0.0976 | -0.1778 | -0.1466 | 0.0468 | 0.0056 | -0.0373 | -0.1192 | -0.0917 |
| **Insulin** | 0.4851 | 0.3403 | 1.0000 | -0.2859 | -0.2628 | -0.2397 | -0.1784 | 0.1878 | -0.0604 | -0.1821 | -0.2908 | -0.2665 |
| **TAM** | -0.3362 | -0.1561 | -0.2859 | 1.0000 | 0.9622 | 0.9313 | 0.7846 | -0.5645 | 0.2043 | 0.5517 | 0.8252 | 0.7727 |
| **TAV** | -0.3353 | -0.0976 | -0.2628 | 0.9622 | 1.0000 | 0.8869 | 0.8019 | -0.6076 | 0.2798 | 0.5916 | 0.7685 | 0.7560 |
| **AIM** | -0.2525 | -0.1778 | -0.2397 | 0.9313 | 0.8869 | 1.0000 | 0.9139 | -0.4802 | 0.1292 | 0.4759 | 0.7869 | 0.7133 |
| **AIV** | -0.2182 | -0.1466 | -0.1784 | 0.7846 | 0.8019 | 0.9139 | 1.0000 | -0.4696 | 0.1788 | 0.4586 | 0.6680 | 0.6311 |
| **SB** | 0.3839 | 0.0468 | 0.1878 | -0.5645 | -0.6076 | -0.4802 | -0.4696 | 1.0000 | -0.8519 | -0.8362 | -0.5739 | -0.7494 |
| **LPA** | -0.3361 | 0.0056 | -0.0604 | 0.2043 | 0.2798 | 0.1292 | 0.1788 | -0.8519 | 1.0000 | 0.5064 | 0.0871 | 0.2916 |
| **MPA** | -0.2407 | -0.0373 | -0.1821 | 0.5517 | 0.5916 | 0.4759 | 0.4586 | -0.8362 | 0.5064 | 1.0000 | 0.6717 | 0.8871 |
| **VPA** | -0.2617 | -0.1192 | -0.2908 | 0.8252 | 0.7685 | 0.7869 | 0.6680 | -0.5739 | 0.0871 | 0.6717 | 1.0000 | 0.9378 |
| **MVPA** | -0.2758 | -0.0917 | -0.2665 | 0.7727 | 0.7560 | 0.7133 | 0.6311 | -0.7494 | 0.2916 | 0.8871 | 0.9378 | 1.0000 |

Based on the Pearson cross-correlation table above, we can deduce the following:

1. Waist and Insulin were moderately positively correlated (0.4851).
2. Waist circumference was weakly positively correlated with Triglycerides levels (0.1966).
3. TAM, TAV, AIM, and AIV were positively correlated with each other, with TAM and TAV having the strongest correlation (0.9622).
4. SB (sedentary behavior) has a strong negative correlation with LPA, MPA, VPA, and MVPA; most notably with LPA (-0.8519) and MVPA (-0.7494). By definition, time spent sedentary directly reduces the time available to spend in the other activity categories, i.e. time spent in light, moderate, vigorous, and combined moderate-to-vigorous physical activity tends to decrease with increased SB.
5. MPA, VPA, and MVPA are positively correlated with each other, with the strongest correlation between VPA and MVPA (0.9378). These categories tend to increase or decrease together.